\documentclass[aps,pre,twocolumn,showpacs]{revtex4-1}  
\usepackage{graphicx}  
\usepackage{dcolumn}   
\usepackage{bm}        
\usepackage{amssymb,amsmath}   
\usepackage{color}

\begin{document}

\title{Coarse-grained patterns in multiplex networks}
\author{Daniel M. Busiello$^{1}$, Timoteo Carletti$^{2}$ and Duccio Fanelli$^{3}$}
\affiliation{$1.$ Department of Physics and Astronomy `G. Galilei' and INFN, Universit\'a di Padova, Via Marzolo 8, 35131 Padova, Italy \\
$2.$ Department of Mathematics and Namur Institute for Complex Systems - naXys, University of Namur, rempart de la Vierge 8, B 5000 Namur, Belgium \\
$3.$ Department of Physics and Astronomy, University of Florence, INFN and CSDC, Via Sansone 1, 50019 Sesto Fiorentino, Florence, Italy}

\date{\today}

\begin{abstract}
A new class of  patterns for multiplex networks is studied, which consists in a collection of different homogeneous states each referred to a distinct layer. The associated stability diagram exhibits a tricritical point, as a function of the inter-layer diffusion coefficients.  The coarse-grained patterns made of alternating homogenous layers, are dynamically selected via non homogeneous perturbations  superposed to  the underlying, globally homogeneous, fixed point and by properly modulating the coupling strength between layers. Furthermore, layer-homogenous fixed points can turn unstable following  a mechanism {\it  \`a la} Turing, instigated by the intra-layer diffusion. This novel class of solutions enriches the spectrum of dynamical phenomena 
as displayed within the variegated realm of multiplex science.
\end{abstract}

\maketitle


Countless systems in Nature exhibit patterns and regularities. Chemistry \cite{murray,zhabo}, biology \cite{suweis,sparsity} and neuroscience \cite{beggs} are just few examples of fields in which a macroscopic order spontaneously emerges from the microscopic interplay between many interacting agents.

A particular subset of processes driving the onset of patterns is represented by reaction-diffusion systems, i.e. systems made of at least two interacting species undergoing spatial diffusion. Introduced in the context of mammals pigmentation by Alan Turing \cite{turing}, from which the celebrated name of Turing patterns, these systems obey an activator-inhibitor dynamics. The diffusion drives an instability by perturbing an homogeneous stable fixed point, under certain conditions. The perturbation grows and, balanced by non linear interactions, leads to spatially inhomogeneous steady states.

Although regular lattices define a suitable framework to model physical reaction-diffusion systems, recently the theory has been extended so as to include  complex networks \cite{nakao}, as the underlying medium where species are bound to diffuse. This approach is motivated by the fact that many real-world systems, ranging from ecology \cite{may} to the brain structure \cite{sporns}, passing through the modeling of social communities \cite{wass}, can be easily schematised by invoking the concept of graph \cite{latora}. In this context, numerous works have revealed a plethora of interesting phenomena, ascribing to the discrete nature of the embedding support a leading role \cite{asymm,cartesian}.

However, the standard approach to network theory is not always able to encode for the high complexity of real-world systems, e.g. the human brain \cite{brain} and the transportation network \cite{kurant,zou}. For this reason, a step forward in the modeling has been made by introducing the concept of multiplex networks, i.e. interconnected multi-layered graph \cite{grow}. A general theory for Turing pattern on multiplex networks has been developed \cite{multiplex,kouvaris}. Interestingly,   diffusion among adjacent layers can enhance or suppress the instability \cite{multiplex}.

\begin{figure}[h]
\centering
\includegraphics[width=\columnwidth]{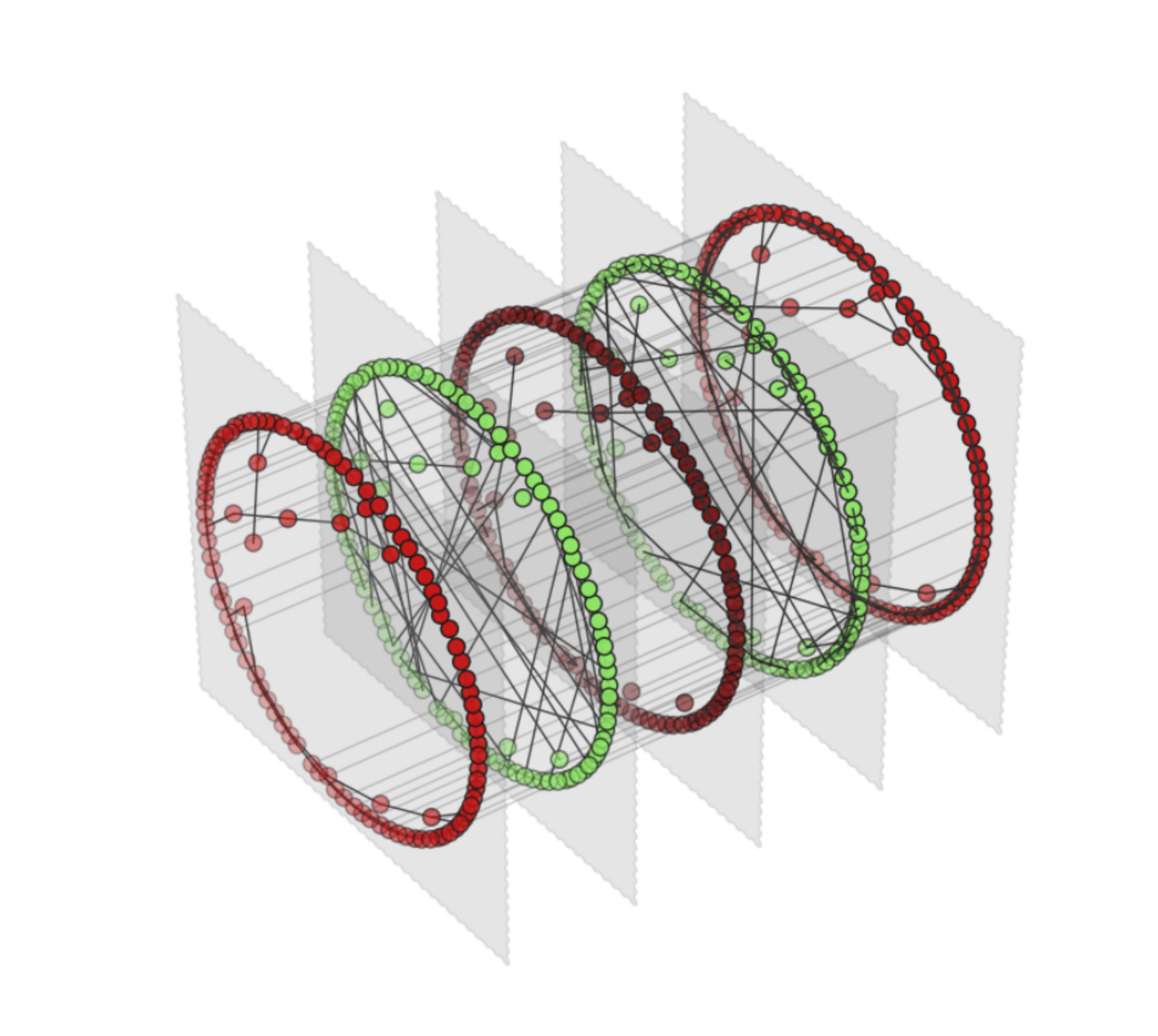}
\caption{Illustrative example of a layer-homogeneous fixed point as obtained for the case of a multiplex network composed by $M=5$ Watt-Strogatz layers  
\cite{watts},  with probability of rewiring $p=0.5$ and average connectivity ranging from $2$ to $5$. Each network is made of $N=100$ nodes. Here, the Brusselator model is assumed, with parameter $b=9$, $c=30$. The diffusion constant are set to the values $D_u^{12} = D_u^{23} = D_u^{34} = D_u^{45} = 1$ and $D_v^{12} = D_v^{23} = D_v^{34} = D_v^{45} = 10$. To facilitate visualization, only $30 \%$ of the links in each layer  and $40 \%$ of the links among layers have been drawn.}
\label{Fig_1}
\end{figure}

In this work we analyse further the zoology of phenomena that can emerge from a reaction-diffusion system defined on multiplex networks. In particular, we focus on a new class of instability driven patterns, veritable attractor of the inspected system, which are homogeneous per layer \cite{tang}, as depicted in Fig. \ref{Fig_1}. These states can turn unstable due to the injection of a non homogeneous perturbation which may resonate with the intra-layer diffusion terms. Each layer of a multiplex network can be also though as an individual node of a corresponding 
coarse-grained graph. In this setting, it is tempting to interpret the novel family of fixed points as coarse-grained patterns, which combines different macro-units so as to reflect
the complexity of a multi-layers arrangement. We shall also prove that such coarse-grained patterns can be dynamically selected following a Turing like instability of the global homogeneous equilibrium, the inter-layer diffusivity acting as the key control parameter.

Let us consider for a sake of simplicity a multiplex composed by two layers, but observe that the model can be readily extended to the case of $M$ independent layers; each layer is constituted by $N$ nodes, and characterised by a $N\times N$ adjacency matrix $A_{ij}^{K}$, where the label $K=1,2$ denotes the layer of pertinence. By definition, $A_{ij}^{K} = 1$ if the nodes $i$ and $j$ are connected in the layer $K$, $A_{ij}^{K} = 0$ otherwise. Let us observe that homologous nodes, i.e. the ''same node'' belonging to different layers, are, by definition, mutually connected.  A two species reaction-diffusion system can hence be cast in the following form \cite{multiplex}:
\begin{gather}
\dot{u}^{K}_i = f(u^{K}_i,v^{K}_i) + D_u^{K}\sum_{j=1}^{N} L_{ij}^{K} u^{K}_j + D_u^{12} \left(u^{K+1}_i-u^{K}_i\right) \nonumber \\
\dot{v}^{K}_i = g(u^{K}_i,v^{K}_i) + D_v^{K}\sum_{j=1}^{N} L_{ij}^{K} v^{K}_j + D_v^{12} \left(v^{K+1}_i-v^{K}_i\right)
\label{multiplex}
\end{gather}
assuming $K=1,2$ and $K+1$ to be $1$ for $K=2$. Here $u^K_i$ and $v^K_i$ stand for the concentrations of the species on the node $i$, as seen in layer $K$. $L_{ij}^{K}$ is the Laplacian matrix associated to the $K$ layer, $L_{ij}^{K} = A_{ij}^{K} - k_i^{K} \delta_{ij}$, where $k_i^{K} = \sum_j A_{ij}^{K}$ refers to the connectivity of node $i$ belonging to layer $K$ \footnote{Notice that $k_i^K$ does not account for inter-layers links.} and $\delta_{ij}$ is the Kroenecker's delta. The matrix $L_{ij}^{K}$ is nothing but the discrete version of a diffusion operator. $D_u^{K}$ (resp. $D_v^{K}$) is the intra-layer diffusion coefficient of species $u$ (resp. $v$); $D_u^{12}$ (resp. $D_v^{12}$) denotes the inter-layer diffusion coefficient associated to species $u$ (resp. $v$). Finally, the non linear functions $f(\cdot,\cdot)$ and $g(\cdot,\cdot)$ encode for the local (on site) rule of interaction between the two considered species. In the following we shall assume that one species acts as an activator, by autocatalytically enhancing its own production, while the other behaves as an inhibitor, contrasting the activator growth.

The model in Eq. \eqref{multiplex} admits two classes of fixed points: (i) the globally homogeneous (GH) fixed points, i.e. $u_i^{K} = \hat{u}$ and $v_i^{K} = \hat{v}$ for all $i=1,...N$ and for all $K$, namely the equilibrium values are independent from the node and the layer; (ii) the layer-homogeneous  (LH) fixed points, defined as $u_i^{K} = \hat{u}^{K}$ and $v_i^{K} = \hat{v}^{K}$ for all $i=1,...N$. Note that we here emphasised the dependence of the equilibrium value on the layer, through the index $K$.

To be concrete, let us consider a specific case study, the so-called Brusselator model for which the local reaction terms are given by $f(u,v) = 1-(b+1)u+cu^2v$ and $g(u,v)=bu-cu^2v$, depending on the parameters $b$ and $c$. A straightforward computation allows one to determine the GH fixed point $\hat{u} = 1$, $\hat{v} = b/c$. Determining the LH fixed point proves more demanding and, to this end, we rely on numerical methods. In the following, the parameters of the model are assigned so that the corresponding GH fixed point is stable to external homogenous perturbation. In other words, ($b,c$) are selected in the region where the a-spatial version of the Brusselator (i.e. the model obtained when setting to zero all diffusion constants in Eqs. (\ref{multiplex})) returns a stable fixed point. 

Note that the GH fixed point depends only on the model parameters $b$ and $c$, since both the intra-layer and the inter-layer diffusions vanishes when all the nodes share the same concentration, independently on the layer that they are bound to occupy. On the other hand,  LH fixed points are also function of the inter-layer diffusion coefficients, $D_u^{12}$ and $D_v^{12}$, but do not depend on the intra-layer diffusion constants, because the species display an identical concentration on each layer, at equilibrium.

\begin{figure*}
\centering
\includegraphics[width=1.5\columnwidth]{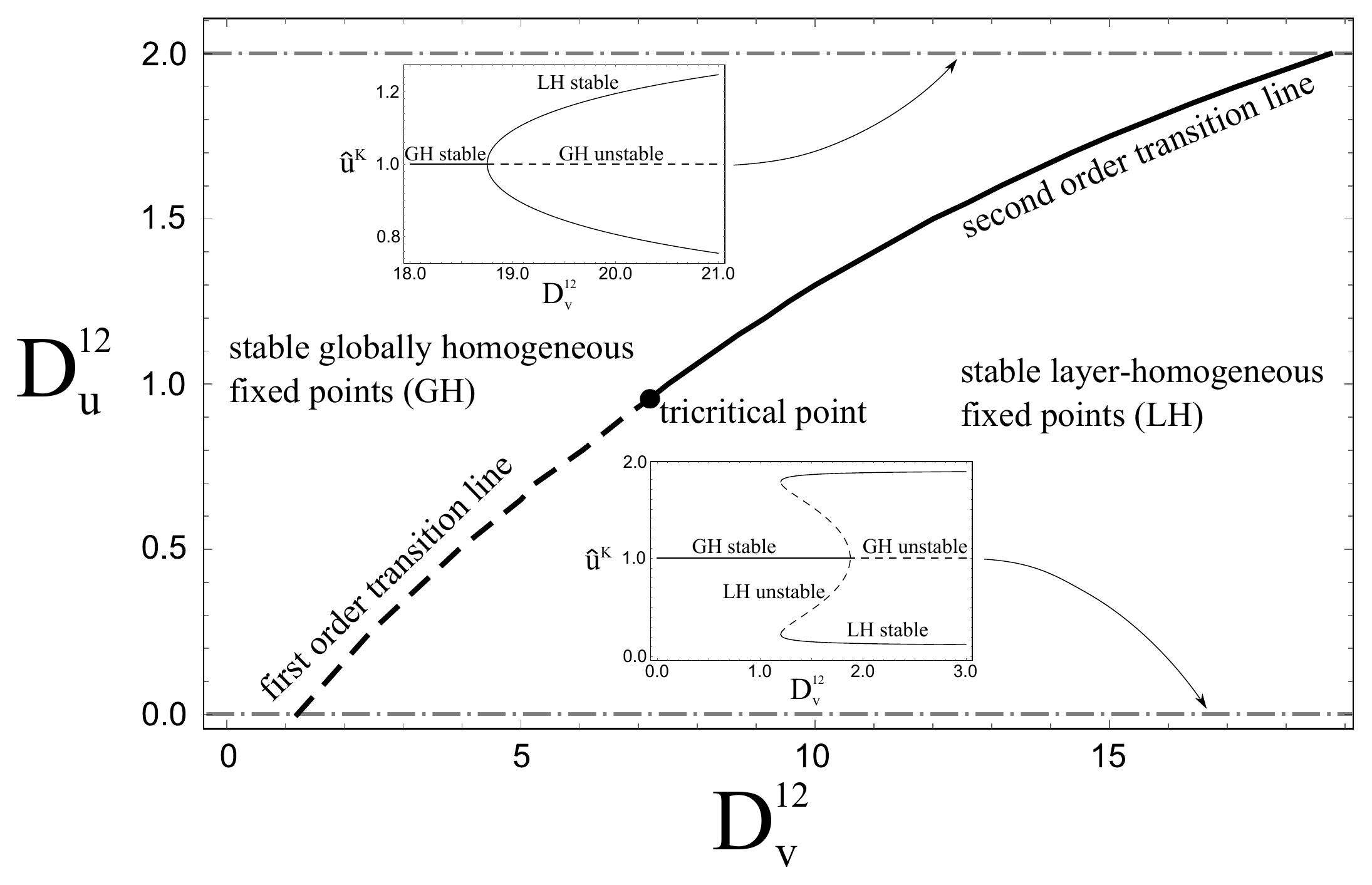}
\caption{\textit{Main} - Phase transitions in the reference plane $(D_u^{12},D_v^{12})$ for a multiplex network composed of two layers. Each layer is made of a Watts-Strogatz network with $N=100$ nodes, probability of rewiring $p=0.5$. The upper Watts-Strogatz network originates from a ring with just nearest-neighbors couplings, while for the other network a third nearest-neighbors lattice is assumed, as the initial skeleton. The curves refer to the Brusselator reaction model with $b=9$ and $c=30$. The dashed line refers to first oder transition, the solid line stands for the second order transition, the circle identifies the position of the tricritical point. In the lower portion of the plane,  LH solutions are stable. Viceversa, in the upper region of the plane, GH fixed points represent the stable equilibria. \textit{Upper inset} - Second order bifurcation diagram: $\hat{u}^K$ is plotted as a function of $D_v^{12}$, for $D_u^{12}=2$ (dash-dotted upper line). GH solutions return an identical value of $\hat{u}^K$, on both layers. GH solutions appear hence as a degenerate single curve. LH  fixed points yield two distinct profiles, each associated to one of the layers of the examined multiplex. \textit{Lower inset} - First order bifurcation diagram: $\hat{u}^K$ is represented as a function of $D_v^{12}$, for $D_u^{12}=0$ (dash-dotted lower line). The qualitative scenario here depicted is robust against modulatining the reaction parameters involved (as e.g. $b$ and $c$), and/or altering the topology of the employed networks. }
\label{Fig1}
\end{figure*}

To study the stability of the above mentioned fixed points, we perform a standard linear stability analysis and monitor the time evolution of a perturbation assumed homogeneous per layer. In formulae, we shall set ${u}^{K}_i = \bar{u} +  \delta{u}^{K}$ and ${v}^{K}_i = \bar{v} +  \delta{v}^{K}$, $\forall i$, where $\bar{u}$ (resp. $\bar{v}$) is either $\hat{u}$ (resp. $\hat{v}$) or $\hat{u}^{K}$ (resp. $\hat{v}^{K}$), and linearize the equations (\ref{multiplex}),  for $\delta{u}^{K}, \delta{v}^{K}$  small. The analysis materializes in an interesting picture, which can be efficaciously summarized in the plane  $(D_u^{12},D_v^{12})$, as reported in  Fig. \ref{Fig1}. The parameters space is partitioned into two regions: in the 
lower portion of the plane LH solutions prove linearly stable. In the upper domain GH fixed points are stable equilibria. The two regions are separated by a transition line which we have determined analytically. The dashed line identifies a first oder transition: by monitoring $\hat{u}^K$, as a function of  $D_v^{12}$, for $D_u^{12}$ frozen to a value that makes the crossing to happen where the transition is predicted discontinuous (horizontal, upper dash-dotted line), one obtains the typical bifurcation diagram as displayed in the rightmost inset of  Fig. \ref{Fig1}. Conversely, when the transition is continuous (horizontal, lower dash-dotted  line) one recovers the usual pitchfork bifurcation, as shown in the rightmost inset enclosed in Fig. \ref{Fig1}. First and second transition lines merge together at a tricritical point, the black circle in Fig. \ref{Fig1}. 

Given the above scenario several interesting questions arise. When operating in the region where the LH fixed point is shown to be stable, can one seed a diffusion driven instability { \`a la Turing}, triggered by a random non homogeneous perturbation? And can one obtain the LH (coarse grained) patterns, as follow a symmetry breaking instability of a GH stable equilibrium? These are the questions that we set to answer in the following. 

Taking inspiration from  \cite{multiplex}, we introduce a small perturbation $(\delta u_i^K, \delta v_i^K)$ to the fixed point $(\hat{u}^K,\hat{v}^K)$ and linearise Eq. \eqref{multiplex} around it:
\begin{equation}
\frac{d}{dt}\left(\begin{array}{ll} {\delta  \textbf{u}} \\ {\delta \textbf{v}} \end{array} \right) = \boldsymbol{\mathcal{J}} \left(\begin{array}{ll} \delta \textbf{u} \\ \delta \textbf{v} \end{array} \right)
\end{equation}
with:
\begin{equation}
\boldsymbol{\mathcal{J}} = \left(\begin{matrix} \boldsymbol{f}_u + \boldsymbol{\mathcal{L}}_u + D_u^{12} \boldsymbol{\mathcal{I}} & \boldsymbol{f}_v \\ \boldsymbol{g}_u & \boldsymbol{g}_v + \boldsymbol{\mathcal{L}}_v + D_v^{12} \boldsymbol{\mathcal{I}} \end{matrix} \right) \nonumber
\end{equation}

where use has been made of the compact notation $\textbf{x} = ( x_1^1,...x_N^1,x_1^2,...x_N^2 )$ for $x=u,v$, and $\boldsymbol{\mathcal{I}} = \left( \begin{matrix} -\boldsymbol{I}_N & \boldsymbol{I}_N \\ \boldsymbol{I}_N & -\boldsymbol{I}_N \end{matrix} \right)$, with $\boldsymbol{I}_{N}$ the $N \times N$ identity matrix. The supra Laplacian \cite{gomez} for species u reads $\boldsymbol{\mathcal{L}}_u = \left(\begin{matrix} D_u^1 \boldsymbol{L}^1 & \boldsymbol{0} \\ \boldsymbol{0} & D_u^2 \boldsymbol{L}^2 \end{matrix} \right)$. Analogously, for  species $v$. Moreover, we have introduced the $4N \times 4N$ matrix 

\begin{equation}
\boldsymbol{f}_u = \left( \begin{matrix} \partial_u f |_{(\hat{u}^1,\hat{v}^1)} \boldsymbol{I}_N & \boldsymbol{0} \\  \boldsymbol{0} & \partial_u f |_{(\hat{u}^2,\hat{v}^2)} \boldsymbol{I}_N \end{matrix} \right) \nonumber
\end{equation}
and, similarly for $\boldsymbol{f}_v$, $\boldsymbol{g}_u$ and $\boldsymbol{g}_v$. These additions  constitute the main difference with respect to the standard Turing theory.

Studying the $4N$ eigenvalues of the matrix $\boldsymbol{\mathcal{J}}$, we can derive the conditions for the dynamical instability which anticipates the onset of the patterns. In fact, if the real part of at least one eigenvalue is positive, the perturbation grows exponentially in the linear regime. Non linear effects become eventually important: they compensate for the linear growth and consequently shape the final non-homogeneous stationary configuration.

From an analytical point of view \cite{multiplex}, one cannot introduce a basis to expand the perturbations which diagonalize the global diffusion operators $\boldsymbol{\mathcal{L}}_u + D_u^{12} \boldsymbol{\mathcal{I}}$ and $\boldsymbol{\mathcal{L}}_v + D_v^{12} \boldsymbol{\mathcal{I}}$. We cannot hence simplify the $4N \times 4N$ eigenvalue problem, by projecting it into a reduced subspace, as it is instead possible when the dynamics is hosted on just one isolated layer  \cite{nakao}. Moreover, the spectrum of $\boldsymbol{\mathcal{J}}$ cannot be exactly related to the spectra of the homologous operators, obtained for the limiting setting when the two layers are formally decoupled ($D_u^{12} = D_v^{12} = 0$).

\begin{figure*}
\centering
\includegraphics[width=1.5\columnwidth]{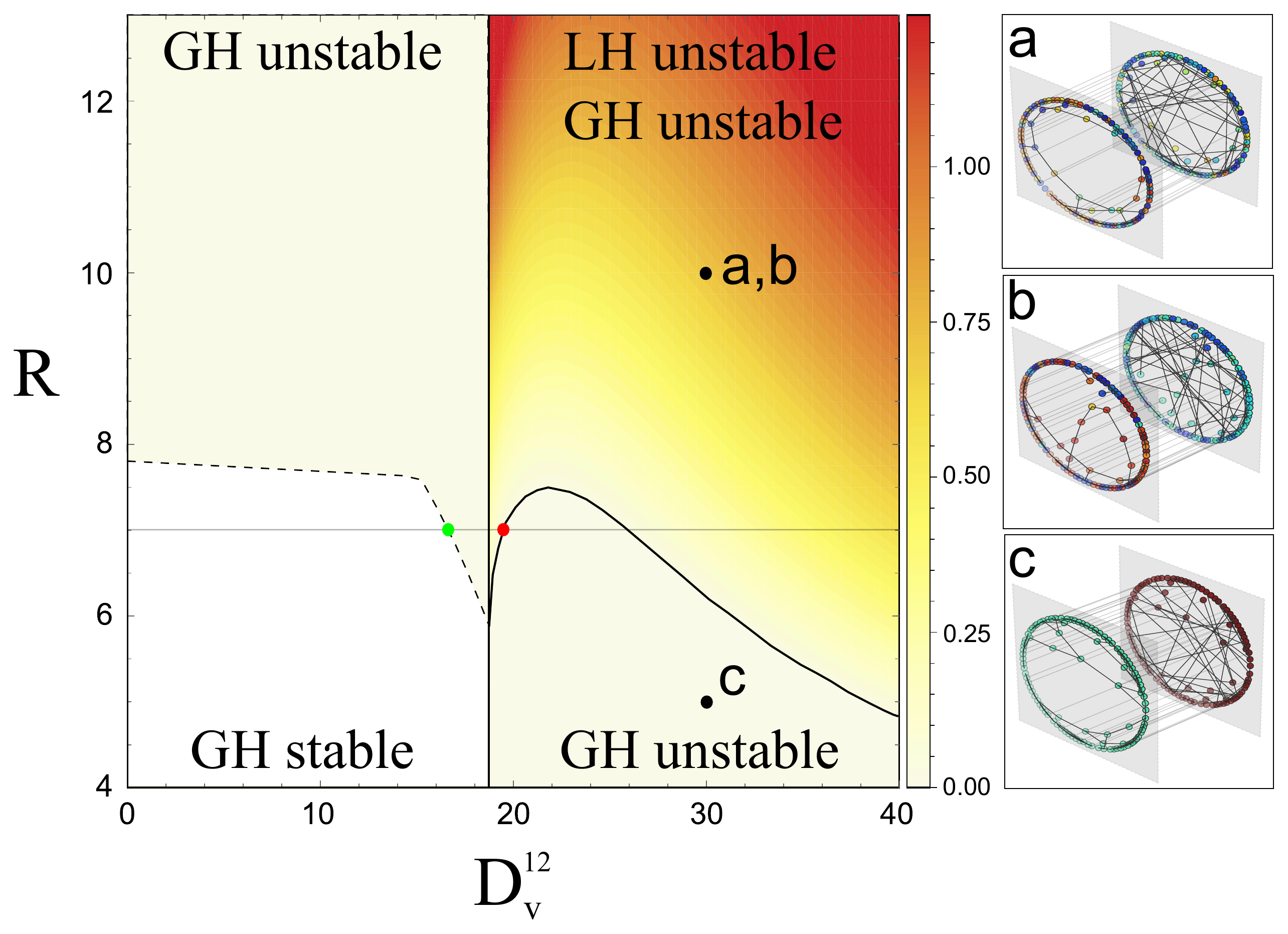}
\caption{The instability domain associated to GH and LH fixed points are studied in the $(D_v^{12},R)$ plane. The Brussellator model is assumed with $b = 9$ and $c = 30$. The inter-layer diffusion coefficient is set to $D_u^{12} = 2$, corresponding to a second order transition in Fig. \ref{Fig1} (similar conclusions can be reached when setting $D_u^{12}$ to a value that yields a first order transition).  The dashed line delineates the upper edge of the GH stability domain (the region of interest is also delimited by the vertical line located at $(D_v^{12})_c$). For $D_v^{12}>(D_v^{12})_c$, the GH is always unstable. Conversely,  LH fixed points turn unstable, upon injection of a non homogeneous perturbation, for a choice of the relevant parameters that position the examined system above the solid line, as obtained via the linear stability analysis.  The dispersion relation, calculated as the largest real part of the eigenvalue of $\boldsymbol{\mathcal{J}}$, evaluated at the LH fixed point is plotted, in the region of LH instability, with an appropriate color code. The patterns obtained upon integration of the system are displayed in the inset: $(a)$ and $(b)$ refer to a choice of the parameters where GH and LH are simultaneously unstable (upper black circle in the main panel). Pattern $(a)$ follows a perturbation imposed on the GH equilibrium, pattern $(b)$ assumes LH as the background equilibrium. Coarse grained patterns can emerge from a GH fixed point, in the region where LH is stable to external perturbation (panel $(c)$). The networks employed have size $N=100$. They have been generated by using the Watts-Strogatz recipe \cite{watts},  with probability of rewiring $p=0.5$, and assuming a first nearest-neighbors ring as an underlying skeleton for the upper layer, and a third nearest-neighbors lattice for the lower. The (green and red) circles in the main figure stand for the transition points between different regimes, as found when cutting at $R=7$. }
\label{Fig2}
\end{figure*}

It is, however, possible to determine numerically the region where the diffusion driven instability is bound to occur, and then integrate the set of governing differential equations, so as to visualize the ensuing patterns. To this aim, we fix the reaction parameters of the inspected model ($b$ and $c$, in the case of the Brusselator model), and one of the inter-layer diffusion coefficients, specifically $D_u^{12}$. When modulating $D_v^{12}$, at fixed $D_u^{12}$, one operates an horizontal cut of Fig. \ref{Fig1}. According to the usual Turing theory, referred to standard mono-layered graphs, the ratio of species' diffusivities matters and ultimately determines the onset of the instability.  In analogy, we here introduce $R^K = D_v^K / D_u^K$, the diffusivities 
ratio as measured on layer $K$. In the following, for the sake of simplicity, we will set $R^1 = R^2 = R$, still limiting the analysis to the minimal choice $K=2$. 
The parameter $R$ keeps track of the intra-layer diffusion, while the inter-layer diffusion (hence the associated ratio) can be solely adjusted by acting on $D_v^{12}$. Several regions are identified, when spanning the parameters in the relevant  plane $(D_v^{12},R)$, see main panel of Fig. \ref{Fig2}. The GH fixed points are stable to external homogeneous perturbations in the (white) region delimited (i) from the above by the dashed solid line, and (ii) from the right by the vertical solid line located at $(D_v^{12})_c \simeq 18$. In the complementary region of the plane, the GH equilibria gets destabilized by injection of a non homogenous disturbance, which is then self-consistently amplified. The solid line depicted for $D_v^{12}>(D_v^{12})_c$, identifies the threshold of instability for the LH fixed point, as determined via the above linear stability analysis. In the domain of LH instability (contained within the region of GH instability), the largest real part of the eigenvalues (also called dispersion relation) of  $\boldsymbol{\mathcal{J}}$, evaluated at the LH fixed point, is represented with an an appropriate color code. Several comments are mandatory at this points. First, LH Turing patterns are instigated by increasing the inter-layer diffusion constant $D_v^{12}$ (or, equivalently, the  ratio $D_v^{12}/D_u^{12}$). Interestingly by tuning $D_v^{12}$ one can drive unstable a system which is otherwise stable, under the classical Turing paradigm applied to GH equilibria, for $D_v^{12}=0$. Further, two alternative pathways can be pursued for pattern generation, in the region where GH and LH  are both unstable. The emerging patterns are annexed as insets $(a)$ and $(b)$ of  Fig. \ref{Fig2}, and display similar qualitative traits. Moreover,  a diffusion driven instability of a GH equilibrium can take the system towards the basin of attraction of a stable LH fixed point, as shown in panel $(c)$ of Fig. \ref{Fig2}.  As anticipated,  it is therefore possible to obtain a LH equilibrium as the dynamical outcome of a genuine Turing instability, acting on a GH fixed point. LH equilibria could be hence interpreted as a special class of Turing patterns. Remark that these latter could be radically simplified at the coarse grained scale, by replacing each layer with a single macro-unit, which bears the very same concentration that happens to be shared by the nodes that define the fine structure of the layer.

To discriminate between distinct classes of ensuing patterns we introduce the following indicator:

\begin{equation}
\Gamma = \frac{1}{N} \sum_{i=1}^N \Theta\left( \left| u_i^1 - u_i^2 \right| \right) - \frac{1}{4N} \sum_{K=1}^2 \sum_{i=1}^N \Theta\left( \left| \sum_{j=1}^N L_{ij}^K u_j^K \right| \right) \nonumber
\end{equation}

where $\Theta(\cdot)$ is the Heaviside function. By evaluating $\Gamma$ at large times, i.e. when the system has eventually reached its asymptotic configuration, one obtains three possible outcomes. When the system is in confined in a globally homogeneous fixed point, $\Gamma = 0$, since both contributions entering the above definition vanish identically. On the contrary, if the system exhibits a disordered pattern, the first term is approximately equal to one, because in general $u_i^1 \neq u_i^2$, and the second term is equal to $1/2$, so yielding $\Gamma \simeq 1/2$. Finally, when the system lands in a layer-homogeneous fixed point, i.e. a coarse-grained pattern, $\Gamma = 1$, since the second term vanishes, being all the concentrations in the same layer equal. A similar indicator could be introduced to monitor the degree of large scale organization of the competing species $v$.

\begin{figure}[h]
\centering
\includegraphics[width=\columnwidth]{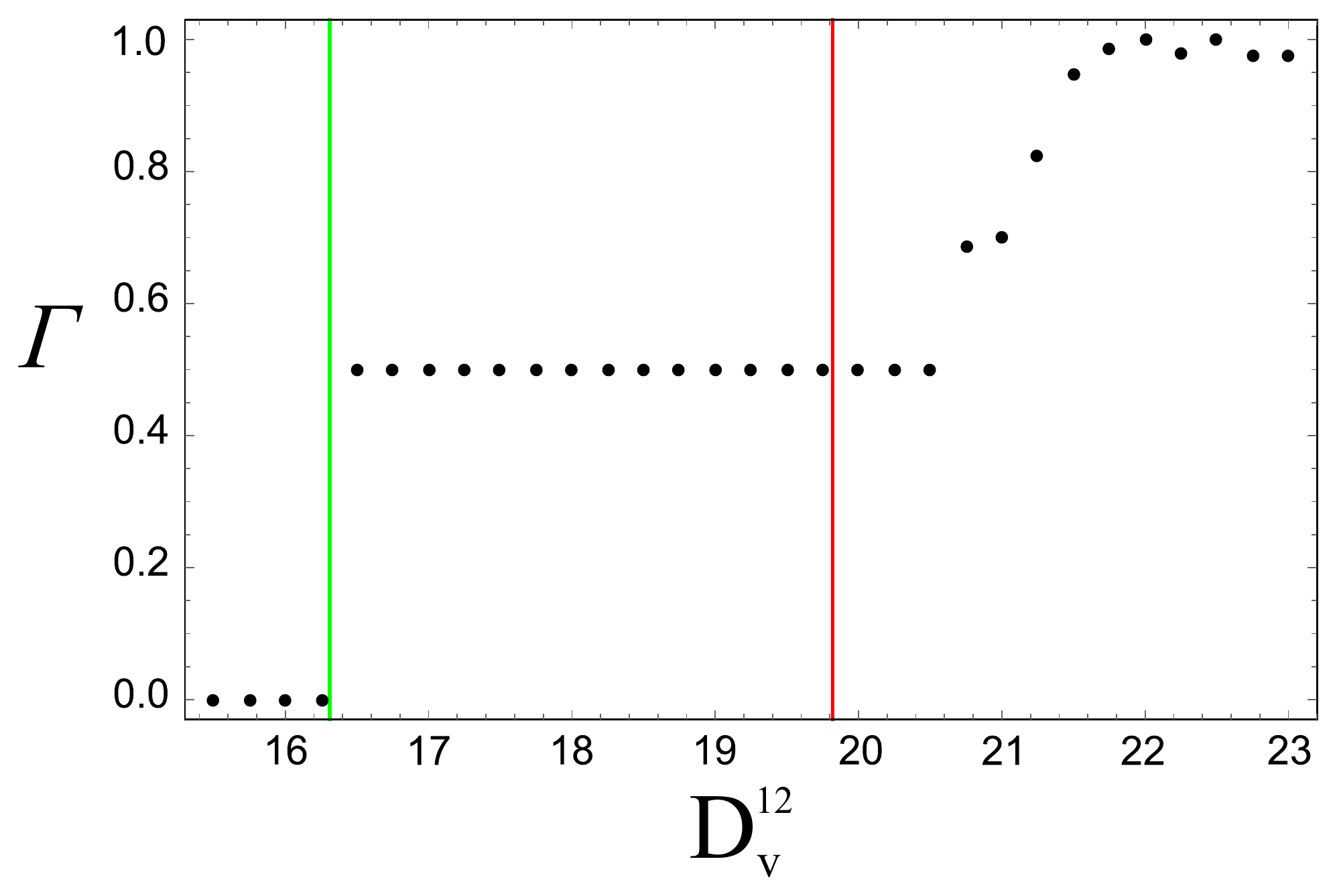}
\caption{$\Gamma$ as a function of the inter-layer diffusion coefficient $D_v^{12}$, as obtained for a Brusselator model with $b=9$ and $c=30$. Here, $D_u^{12} = 0$ and $R=7$ (see horizontal line in Fig. \ref{Fig2}). The multiplex network is composed of two layers, generated as explained in the caption of Fig. \ref{Fig2}. Each reported entry refers to averaging over $20$ simulations, as the emergence of a LH attractor is a probabilistic event. The green vertical line is traced for a value of $D_v^{12}$ that marks the instability of GH fixed point, when moving at fixed $R=7$ (green point in Fig.  \ref{Fig2}). The red vertical line sets the beginning of the region where LH solutions are stable attractors of the dynamics. Representative examples of the obtained patterns are depicted in the insets of Fig. \ref{Fig2}.}
\label{Fig3}
\end{figure}

In Fig. \ref{Fig3} we report the value of $\Gamma$, as a function of the inter-layer diffusion coefficient $D_v^{12}$. The emerging patterns, classified in terms of $\Gamma$, are obtained by perturbing a GH fixed point, assumed stable in absence of diffusion. The results displayed Fig. \ref{Fig3} are computed by processing the patterns recorded via direct numerical integration of the equations (\ref{multiplex}), and after averaging over different realizations of dynamics.  The results refer  $D_u^{12} = 2$ (second order transition), being the other parameters frozen to the values declared in the caption of Fig. \ref{Fig1} (qualitatively similar results are obtained when setting $D_u^{12}$ to a value which corresponds to a first order transition). By eye inspection, it is immediate to conclude that the system exhibits disordered or coarse-grained patterns, depending on the value of the coupling strength between layers. The regions when different patterns are found, organize in adjacent blocks, as function of the control parameter $D_v^{12}$. More specifically, for modest values of $D_v^{12}$, standard Turing patterns take place. By increasing $D_v^{12}$, coarse-grained patterns are instead established. The vertical (green and red) lines are respectively drawn in correspondence of the critical values of $D_v^{12}$, as identified in Fig. \ref{Fig2}, see (red and green) circles. These latter provide a satisfying theoretical interpretation for the observed transitions.

Summing up,  we have here developed a theory of pattern formation for a two species reaction diffusion system on a two-layers multiplex network around a novel class of (layer-homogeneous) fixed points.  The methodology can be readily generalized to the case of $s$ interacting species upon $K$ independent layers. Interestingly,  the layer-homogeneous fixed points can be seen as coarse-grained patterns in a network of networks where each layer is replaced by a virtual super-node, bearing the concentration displayed by individual micro-nodes, belonging to the layer itself. This novel class of patterns can be dynamically selected by perturbing the globally homogeneous fixed point and assigning the coupling strength between different layers, to fall within a specific window. The inter-layer diffusion can promote the onset of patterns formation. This scenario has been tested with reference to the Brusselator model, assumed as a paradigmatic representative, but it holds in general, irrespectively of the specific reaction rules implemented. Different kind of patterns co-exist in systems defined on a stratified network and the diffusion act as trigger to resolve  hidden microscopic structures. The potential interest of this observation embraces a large set of applications, ranging from epidemic spreading to ecological interactions, passing through the study of pattern formation in neural networks.

\section*{Acknowledgments}
The work of T.C. presents research results of the Belgian Network DYSCO 
(Dynamical Systems, Control, and Optimization), funded by the Interuniversity 
Attraction Poles Programme, initiated by the Belgian State, Science Policy Office. 
The scientific responsibility rests with its author(s).  D.F. acknowledges financial support 
from H2020-MSCA-ITN-2015 project COSMOS  642563.


\begin{thebibliography}{10}

\bibitem{murray}
J.~D. Murray, {\em Mathematical Biology. II Spatial Models and Biomedical
  Applications $\{$Interdisciplinary Applied Mathematics V. 18$\}$}.
\newblock Springer-Verlag New York Incorporated, 2001.

\bibitem{zhabo}
A.~M. Zhabotinsky, M.~Dolnik, and I.~R. Epstein, ``Pattern formation arising
  from wave instability in a simple reaction-diffusion system,'' {\em The
  Journal of chemical physics}, vol.~103, no.~23, pp.~10306--10314, 1995.

\bibitem{suweis}
S.~Suweis, F.~Simini, J.~R. Banavar, and A.~Maritan, ``Emergence of structural
  and dynamical properties of ecological mutualistic networks,'' {\em Nature},
  vol.~500, no.~7463, pp.~449--452, 2013.

\bibitem{sparsity}
D.~M. Busiello, S.~Suweis, J.~Hidalgo, and A.~Maritan, ``Explorability and the
  origin of network sparsity in living systems,'' {\em Scientific reports},
  vol.~7, no.~1, p.~12323, 2017.

\bibitem{beggs}
J.~M. Beggs and D.~Plenz, ``Neuronal avalanches in neocortical circuits,'' {\em
  Journal of neuroscience}, vol.~23, no.~35, pp.~11167--11177, 2003.

\bibitem{turing}
A.~M. Turing, ``The chemical basis of morphogenesis,'' {\em Philosophical
  Transactions of the Royal Society of London. Series B, Biological Sciences},
  vol.~237, no.~641, pp.~37--72, 1952.

\bibitem{nakao}
H.~Nakao and A.~S. Mikhailov, ``Turing patterns in network-organized
  activator-inhibitor systems,'' {\em Nature Physics}, vol.~6, no.~7,
  pp.~544--550, 2010.

\bibitem{may}
R.~M. May, {\em Stability and complexity in model ecosystems}, vol.~6.
\newblock Princeton university press, 2001.

\bibitem{sporns}
O.~Sporns, D.~R. Chialvo, M.~Kaiser, and C.~C. Hilgetag, ``Organization,
  development and function of complex brain networks,'' {\em Trends in
  cognitive sciences}, vol.~8, no.~9, pp.~418--425, 2004.

\bibitem{wass}
S.~Wasserman and K.~Faust, {\em Social network analysis: Methods and
  applications}, vol.~8.
\newblock Cambridge university press, 1994.

\bibitem{latora}
S.~Boccaletti, V.~Latora, Y.~Moreno, M.~Chavez, and D.-U. Hwang, ``Complex
  networks: Structure and dynamics,'' {\em Physics reports}, vol.~424, no.~4,
  pp.~175--308, 2006.

\bibitem{asymm}
M.~Asllani, J.~D. Challenger, F.~S. Pavone, L.~Sacconi, and D.~Fanelli, ``The
  theory of pattern formation on directed networks,'' {\em Nature
  communications}, vol.~5, p.~4517, 2014.

\bibitem{cartesian}
M.~Asllani, D.~M. Busiello, T.~Carletti, D.~Fanelli, and G.~Planchon, ``Turing
  instabilities on cartesian product networks,'' {\em Scientific reports},
  vol.~5, 2015.

\bibitem{brain}
E.~Bullmore and O.~Sporns, ``Complex brain networks: graph theoretical analysis
  of structural and functional systems,'' {\em Nature Reviews Neuroscience},
  vol.~10, no.~3, pp.~186--198, 2009.

\bibitem{kurant}
M.~Kurant and P.~Thiran, ``Layered complex networks,'' {\em Physical review
  letters}, vol.~96, no.~13, p.~138701, 2006.

\bibitem{zou}
S.-R. Zou, T.~Zhou, A.-F. Liu, X.-L. Xu, and D.-R. He, ``Topological relation
  of layered complex networks,'' {\em Physics Letters A}, vol.~374, no.~43,
  pp.~4406--4410, 2010.

\bibitem{grow}
V.~Nicosia, G.~Bianconi, V.~Latora, and M.~Barthelemy, ``Growing multiplex
  networks,'' {\em Physical review letters}, vol.~111, no.~5, p.~058701, 2013.

\bibitem{multiplex}
M.~Asllani, D.~M. Busiello, T.~Carletti, D.~Fanelli, and G.~Planchon, ``Turing
  patterns in multiplex networks,'' {\em Physical Review E}, vol.~90, no.~4,
  p.~042814, 2014.

\bibitem{kouvaris}
N.~E. Kouvaris, S.~Hata, and A.~D{\'\i}az-Guilera, ``Pattern formation in
  multiplex networks,'' {\em Scientific reports}, vol.~5, p.~srep10840, 2015.

\bibitem{tang}
L.~Tang, X.~Wu, J.~L{\"u}, J.-a. Lu, and R.~M. D'Souza, ``Master stability
  functions for multiplex networks,'' {\em arXiv preprint arXiv:1611.09110},
  2016.

\bibitem{watts}
D.~J. Watts and S.~H. Strogatz, ``Collective dynamics of
  ‘small-world’networks,'' {\em nature}, vol.~393, no.~6684, pp.~440--442,
  1998.

\bibitem{Note1}
Notice that $k_i^K$ does not account for inter-layers links.

\bibitem{gomez}
S.~Gomez, A.~Diaz-Guilera, J.~Gomez-Gardenes, C.~J. Perez-Vicente, Y.~Moreno,
  and A.~Arenas, ``Diffusion dynamics on multiplex networks,'' {\em Physical
  review letters}, vol.~110, no.~2, p.~028701, 2013.

\end{thebibliography}
\end{document}